\documentclass[preprint,12pt]{elsarticle}

\usepackage{amssymb}

\usepackage[nodots,nocompress]{numcompress}

\biboptions{super}

\journal{Nuclear Physics B}

\begin{document}

\begin{frontmatter}


\ead{longo@na.infn.it}

\title{Astroinformatics, data mining and the future of astronomical research}


\author[a]{Massimo Brescia}
\address[a]{INAF - Astronomical Obs. of Capodimonte, via Moiariello 16, I-80131, Napoli, Italy}
\author[b,c]{Giuseppe Longo}
\address[b]{Dept. of Physics, University Federico II, via Cintia 6, 80126, Napoli, Italy}
\address[c]{visiting associate - Dept. of Astronomy, Caltech, Pasadena, USA}
\begin{abstract}
Astronomy, as many other scientific disciplines, is facing a true data deluge which is bound to change 
both the praxis and the methodology of every day research work. The emerging field of astroinformatics, 
while on the one end appears crucial to face the technological challenges, on the other is opening new
exciting perspectives for new astronomical discoveries through the implementation of advanced data mining 
procedures. 
The complexity of astronomical data and the variety of scientific problems, however, call for 
innovative algorithms and methods as well as for an extreme usage of ICT technologies.

\end{abstract}


\end{frontmatter}


\section{Introduction}\label{introduction}
Astroinformatics is an emerging discipline with rather ill defined boundaries encompassing all those   
aspects of computing and ICT technologies which are crucial to deal with the data streams and data sets
produced by a new generation of instruments, sensors and computer simulations. 
There is at least one good reason to assess Astroinformatics as a new discipline rather 
than as a simple amalgam of problems and methodologies collected from other fields: 
by facing problems which differ from those encouteres in any other discipline 
both in type and in size, Astroinformatics is triggering a true methodological shift
within the astronomical community. This can be better understood by taking into account 
data quantity and data complexity.

The first one, data quantity, is obvious: the new generation of telescopes, instruments and sensors 
produces or will do so, data streams in the many terabytes or even petabytes regime which cannot be 
effectively explored with traditional tools, both hardware and software. 
Such data volumes imply that most processing, reduction and analysis, needs to be performed in a semi 
or fully automatic fashion using computing infrastructures, methods and tools which push to the limits 
all existing and even foreseeable ICT technologies. 

Much more challenging is the ``data complexity" aspect. One way to exemplify some of its 
deeper implications is through the concept of ``astronomical parameter space" which, 
at least in its simplest form (i.e. without taking into account the role played by errors and 
uncertainties, and the fact that many of its dimensions are not independent) is almost trivial.
The parameter space can be seen as the $N$ dimensionality manifold ($\aleph^N\subset \Re^N$) 
defined by the astronomical observables. 
Each observations can be expressed as a string of numerical measures -e.g. Right ascension, declination, 
epoch, flux in a given band, morphological type, etc.. - 
which projects in $\aleph^N$ as a point or as a hyperplane or hypervolume of lower 
dimensionality. 
In order to clarify this idea let us assume $N=15$. 
An observation measuring all 15 parameters would be a point in $\aleph^N$; if only 14 parameters 
are measured, it identifies
a line in $\aleph^N$; if only 13,  an hypersurface, and so on.
As first pointed out by Martin Harwit \cite{1,2}, the entire history of astronomical discoveries can be re-read as 
the history of an ever evolving parameter space where new dimensions were added as new technologies opened new observational 
windows and/or improved the sampling of previous ones. 
For instance, quasars were disentangled from stars when the radio flux dimension was added; 
pulsars when the sampling of the radio flux was increased down to a few milliseconds, etc.  

As long as the parameter space can be projected on two or three dimensions, it can be visualised and
a trained human eye can easily spot trends or patterns (i.e, for instance, empirical laws or peculiar classes 
of objects). But what, if these correlations take place among more than three independent variables and are 
ill defined in any 3-D subspace? They would almost surely escape detection.
In the past this was not an issue, but now that the federation of most existing and future astronomical data bases 
and data sets in the cyberinfrastructure called Virtual Observatory\cite{3} (VObs) has deeply and forever altered 
the astronomer's playground, things have changed.  
The fact that most astronomical data is or will soon be accessible with a few clicks of the mouse will surely
prompt ever more complex questions requiring complex data queries in a distributed federation of data repositories.
\section{The present}

One crucial aspect is the fact that most DM algorithms are not robust against missing data and 
cannot effectively deal with upper limits. 
In fields other than astronomy (e.g. market analysis and many if not all bioinformatics applications) 
this is only a minor problem since the data set to be mined 
can be cleaned of all those records having incomplete or missing information and the 
results can be afterwords generalized. 
To be more explicit: if in an anagraphic record for one citizen the field "`age"' is missing, it means that 
the data has not been collected and not that the citizen has no age; if an object is missing
a magnitude in a specific photometric band it may either mean that it has not been observed (as in the previous case)
or, and this is much more interesting, that it is so faint that it cannot be detected in that band.  
In the first case, the missing information can be easily recovered if a proper data model exists 
and the machine learning literature offers a variety of models which can reconstruct the missing 
information. 
A similar approach has also been adopted by the astronomical community in some simple applications such 
as, for instance, star/galaxy separation \cite{4} or evaluation of photometric 
redshifts \cite{5} with neural networks.  
In the second case, which includes many of the most interesting astronomical applications 
such as, for instance, the search for obscured or high redshift quasars, the crucial information is in the missing data 
themselves and this information cannot be reconstructed.

It is not clear to the writers whether the problem can be solved by adapting existing methods or 
rather requires the the implementation of a new geenration of ML learning 
methods. 
The first case seems however unlikely since it would require methods based on adaptive metrics
capable to work equally well on points, hypersurfaces and hypervolumes of different dimensionality.

A second serious threat is posed by scalability of existing algorithms and methods. 
It is in fact well known that most if not all existing Data mining methods scale badly with both 
increasing number of records and/or of features (i.e. input variables).
When working on massive data sets this problem is usually circumvented by extracting subsets of 
data, performing the training and validation of the methods on these decimated data sets and 
then extrapolating the results to the whole set. 
This approach obviously introduces biases which are often difficult to control but, more important, 
there are at least two reasons why it cannot be a viable solution neither on the long term nor 
if data mining has to be adopted by the community at large.

First of all, because the Data Mining praxis requires, for a given problem, a lenghty fine tuning procedure 
which implies ten's and sometimes hundreds of experiments to be performed in order to identify the optimal 
method or, within the same method, the optimal architecture or combination of parameters.
In the case of MDS, even the decimated data would still pose serious computational challenges 
which would be unmatchable by most users.
Second, astronomical data mining is extremely heterogeneous and the MDS need to be accessed and used by a 
large community of thousands of different users each with different goals, scientific interests and methods.
It is therefore unthinkable to move such data volumes across the network from the distributed data 
repositories to a myriad of different users.

\section{DAME - Data Mining and Exploration}

In order to render user friendly and effective DM on astronomical MDS, there are therefore three types of problems 
which need to be solved: to provide the users with an easier access to both methods and computing power;
to identify and implement better and faster algorithms esploiting whereas possible the HPC paradigm; 
to minimize or even reduce the data transfer by moving the programs rather than the data.
The first problem finds its natural solution in the usage of web applications, where the user interacts with the data 
mining framework via a browser and the computing infrastructure is completely transparent to him.  

The DAME (Data Mining  and Exploration) \cite{6} program is based on a platform which allows the scientific community to 
perform data mining and exploration experiments on massive data sets, by using a simple web browser. 
By means of state of the art Web 2.0 technologies (for instance web applications and services), DAME 
offers several tools which can be seen as working environments where to choose data analysis functionalities 
such as clustering, classification, regression, feature extraction etc., together with models and algorithms, 
all derived from the machine learning paradigms.
The user can setup, configure and execute experiments on his own data on top of a virtualized computing infrastructure, 
without the need to install any software on his local machine. 
Furthermore the DAME infrastructure offers the possibility to extend the original library of available tools, 
by allowing the end users to plug-in and execute their own codes in a simple way, by uploading the programs 
without any restriction about the native programming language, and automatically installing them through a simple 
guided interactive procedure.  
Moreover, the DAME platform offers a variety of computing facilities, organized as a cloud of 
versatile architectures, from the single multi-core processor to a grid farm, 
automatically assigned at runtime to the user task, depending on the specific problem, as well as on the 
computing and storage requirements.

\section{A possible future}

There are at least two reasons not to move data sets over the network from their original repositories to the 
user's computing infrastructures. First of all the fact that the transfer could be impossible due to the
available bandwidth and, second, because there could be restrictive policies to data access.
In these cases, the problem is to move the data mining toolsets to the data centers. 
Current strategies, under investigation in some communities such as the VObs, are based on implementing 
web based protocols for application interoperability (e.g. the Web Samp Connector) but this is only a 
partial  solutions since it is still required to exchange data over the web between application sites. 
The DAME collaboration has recently proposed a different approach, which foresees a standardized web application 
repository cloud named "Hydra Lernaen" -- from the name of the ancient snake-like monster with many independent but equal heads.

It consists in Web Application Repositories (WAR) of data mining model and tool packages, to be installed 
and deployed in a generic data warehouse. Different WARs may differ in terms of available models since 
any hosting data center might require specific kinds of data mining and analysis tools. 
If the WARs are structured around a pre-designed set of standards which completely describe their interaction 
with the external environment and application plugin and execution  procedures, two generic data warehouses can 
exchange algorithms and tool packages on demand.
On a specific request the mechanism starts a very simple automatic procedure which moves applications, 
organized under the form of small packages (some MB in the worst case), through the Web from a WAR source to a 
WAR destination, install them and makes the receiving WAR able to execute the imported model on local data. 
More refinements of the above mechanism can be introduced at the design phase, such as for instance to expose, 
by each WAR, a public list of available models, in order to inform other sites about services which could be imported.

Such strategy requires a standardized design approach, in order to provide a suitable set of 
standards and common rules to build and codify the internal structure of WARs and of the data mining applications themselves
(for example any kind of rules like PMML, Predictive Model Markup Language). 
These standards should be designed to maintain and preserve the compliance with data representation 
rules and protocols already defined and currently operative in a particular scientific community 
(such as the VObs in Astronomy). 
Any data Center could implement a suitable computing infrastructure hosting the WAR and thus become a sort of mirror site 
of a  world-wide cross-sharing network of data mining application repository in which it could be engaged a virtuous mechanism of a distributed multi-disciplinary data mining infrastructure, able to deal with heterogeneous or specialized exploration of MDS. 
Such approach seems the only effective way to preserve data ownership and privacy policy, to enhance the e-science community interoperability and to overcome the problems posed by the present and future tsunami of data.





\bibliographystyle{model6-num-names}



\end{document}